\documentclass[%
superscriptaddress,
preprint,
nofootinbib,
 amsmath,amssymb,
 aps,
 prd,
]{revtex4-1}

\usepackage{natbib}

\usepackage{graphicx}
\usepackage{dcolumn}
\usepackage{bm}

\usepackage{epsf}
\usepackage{amsmath,amssymb}
\usepackage{slashed}
\usepackage{graphicx}
\usepackage{subfig}

\usepackage[hidelinks]{hyperref}
\usepackage{comment}

\usepackage{url}

\usepackage{tikz}
\usetikzlibrary{quantikz}
\usepackage{physics}

\begin{document}
\title{Background Suppression in Quantum Sensing of Dark Matter via Collective Entangled-State Projection}

\author{Shion Chen}
\affiliation{Department of Physics, Graduate School of Science, Kyoto University, Kyoto 606-8502, Japan}

\author{Hajime Fukuda}
\email{hfukuda@hep-th.phys.s.u-tokyo.ac.jp}
\affiliation{Department of Physics, The University of Tokyo, Tokyo 113-0033, Japan}

\author{Yutaro Iiyama}
\affiliation{International Center for Elementary Particle Physics (ICEPP),
The University of Tokyo, 7-3-1 Hongo, Bunkyo-ku, Tokyo 113-0033, Japan}

\author{Yuya Mino}
\affiliation{International Center for Elementary Particle Physics (ICEPP),
The University of Tokyo, 7-3-1 Hongo, Bunkyo-ku, Tokyo 113-0033, Japan}

\author{Takeo Moroi}
\email{moroi@hep-th.phys.s.u-tokyo.ac.jp}
\affiliation{Department of Physics, The University of Tokyo, Tokyo 113-0033, Japan}

\author{Mikio Nakahara}
\affiliation{IQM Quantum Computers, Keilaranta 19, Espoo, 02150, Finland}

\author{Tatsumi Nitta}
\affiliation{QUP (WPI), KEK, Oho 1-1, Tsukuba, Ibaraki 305-0801, Japan}

\author{Thanaporn Sichanugrist}
\email{thanaporn@hep-th.phys.s.u-tokyo.ac.jp}
\affiliation{Department of Physics, The University of Tokyo, Tokyo 113-0033, Japan}

\begin{abstract}
We show that measuring dark matter signal by projecting quantum sensors in the collective excited state can highly suppress the non-collective noise background, hence improving the sensitivity significantly.
We trace the evolution of the sensors' state in the presence of both dark matter effect and sensors' decoherence effects, optimizing the protocol execution time, and show that the suppression of background by a factor equal to the number of sensors is possible.
This method does not require the entanglement of sensors during the signal accumulation time, hence circumventing the difficulty of maintaining the lifetime of the entangled state that is present in other enhancement proposals. This protocol is also general regarding the type of qubit sensors.
\end{abstract}

\maketitle

\section{Introduction}
The development of quantum technology and quantum sensing~\cite{Degen:2016pxo,Chou:2023hcc} opens a new possibility for solving one of the decades-old mysteries, the identity of dark matter (DM)~\cite{Zwicky:1937zza,Planck:2018vyg,ParticleDataGroup:2024cfk}. Specifically, quantum bit (qubit) sensors such as superconducting qubits~\cite{Dixit:2020ymh,Chen:2022quj,Chen:2023swh,Agrawal:2023umy,Braggio:2024xed,Chen:2024aya,Dong:2025mdk}, nitrogen-vacancy centers in diamonds~\cite{Chigusa:2023roq,Chigusa:2024psk}, ion traps~\cite{Ito:2023zhp}, and more~\cite{Fan:2022uwu,Engelhardt:2023qjf,Moretti:2024xel,Chigusa:2025rqs}, are shown to be good sensors for wave-like DM direct detection (for review of wave-like DM, see Refs.~\cite{Arias:2012az,Hui:2021tkt}). In particular, the qubit sensors can interact with DM and be excited as an observable signal.
The favorable characteristics of quantum sensors are the state controllability~\cite{Li:2023cla,Li:2024xbj}, precise readout~\cite{Chen:2022frn}, and frequency scan ability~\cite{Koch:2007hay}; besides, their quantum nature~\cite{Lamoreaux:2013koa,Chen:2023swh,Ito:2023zhp,Huang:2024enq,Fukuda:2025zcf} also shows their potential to enhance the signal-to-noise ratio beyond the ability of the classical approach in the setup of DM detection.

The general issues regarding the DM detection are the feeble signal amplitude of DM and the presence of background mimicking the signal, closely tied together.
Since the DM signal itself is extremely feeble, the background is problematic and needs to be considered carefully as well. 
One direction addressing the weak signal amplitude is to enhance the DM signal by quantum-coherently accumulating it with entangled sensors to surpass the background noise~\cite{Chen:2023swh,Ito:2023zhp}.
While this is a powerful way to drastically improve the sensitivity, the full potential requires maintaining the stability of the highly entangled state of the sensors during the signal accumulation time~\cite{Huelga:1997mw}.
Here, we instead focus on the direction to reduce the noise background of the DM signal with quantum sensors and their quantum state manipulation.

One promising way to deal with the background is to take advantage of the fact that wave-like DM interacts with multiple sensors collectively, in contrast to the independent noise background.
At first sight, one may rely on measuring the signal classically at each sensor as a function of time and analyzing their correlation, aiming to reduce the independent random noise while preserving the information of the signal. 
However, in the case of an extremely weak signal as a DM signal, since the signal obtained from each sensor is severely buried by the noise before the signal processing, the correlation analysis is known to become inefficient~\cite{Lamoreaux:2013koa,Degen:2016pxo,Huelga:1997mw,Fukuda:2025zcf}. 
This motivates us to consider measuring the signal with the quantum protocol to directly suppress independent noise at the moment of the measurement, relying on the quantum state manipulation or entanglement between qubit sensors. 

In this paper, we demonstrate that the measurement of the DM signal by the projection of sensors' state to the collective excited state called the $W$ state~\cite{Dur:2000zz}  (the symmetrically superposition between qubits where only one qubit is excited), can significantly suppress the background mimicking the DM signal. 
The key idea is that while DM influences qubit sensors correlatedly and contributes directly to the collective excitation, independent noises instead primarily drive qubits into other subspaces. By carefully selecting the subspace for projection, the noise contribution can therefore be substantially mitigated (see also Ref.~\cite{Shu:2024nmc}).
In this paper, we also focus on more quantitative features 
of the protocol, clarifying the potential and limitations of the protocol both analytically and numerically. We employ the Lindblad equation formalism~\cite{Lindblad:1975ef,Gorini:1975nb}, tracing the sensors' evolution in the presence of both DM signal and sensor noise effects. 
For the noise effects, we consider the generalized amplitude damping and dephasing; i.e., we take into account not only the background excitation but also the deexcitation and dephasing effects.
We optimize the measurement time in the protocol and demonstrate that, without the initial entangled state preparation required, the proposed method can reduce the background excitation significantly by a factor $L$ equal to the number of sensors, compared with the protocol that separately measures qubits and simply counts the total excitation number due to the effect of DM.  
On the other hand, interestingly, we find that we cannot suppress the background noise arbitrarily however large the number of sensors is. This is because the $W$ state can pick up only one excitation; once the number of sensors is large enough, the excitation noise starts to reduce the signals as well, rather than just to mimic the DM signal, by exciting other qubits than the one excited by DM.
The protocol is general with respect to the type of quantum sensors, from various types of qubits to
resonant cavities\,\cite{Fukuda:2025zcf}, provided that the state manipulation of the qubits and the circuit for projecting the state into $W$ state are executable.

The paper is constructed as follows. We first explain the setup model of a qubit in the presence of the DM effect, and define the sensitivity for the DM signal measurement in Sec. 2. Then, we evaluate the sensitivity that can be obtained by separate measurement and $W$ state measurement in Sec. 3 and Sec. 4, respectively. We discuss and conclude results in Sec. 5.

\section{Dark matter and noise model}
In this section, we explain the model of the DM signal and noise effects of qubit sensors and derive the standard deviation of the estimator of the interaction parameter to quantify the sensitivity for the DM signal measurement.
As the quantum sensor, we focus on qubit sensors directly interacting with wave-like DM, such as superconducting transmon qubits (for details, see Ref.~\cite{Chen:2022quj}). However, the discussion in this section is general and can be applied to other types of qubit sensors as well.
For the noise effects, we take Markovian noise into account. To be specific, we consider the background excitation, amplitude damping, and dephasing effects of qubits. 
We assume all noises are local and uncorrelated across sensors. We note that while the protocol we propose in this paper efficiently suppresses uncorrelated background noise, correlated noise acting across multiple sensors might also be present in realistic settings and is not suppressed by our method.

First, we introduce the model of qubit sensors and the DM signal without noise effects.
We consider $L$ qubit sensors of the same frequency $\omega$ interacting with DM through Pauli-$X$ interaction. The Hamiltonian of the system is given by
\begin{align}
    H&= H_0 + H_1, \label{eq:Hsetup} \\
    H_0 &=     -\frac{\omega}{2} \sum^L_{i=1} \sigma^Z_i, \\
    H_1 &= +2 \epsilon \sum_{i=1}^L \sigma^X_i\cos (m t-\varphi),
\end{align}
where $H_0$ is the free Hamiltonian of qubits and $H_1$ denotes the interaction of qubits with DM; $m$ is the mass of DM determining the DM signal frequency, $\varphi$ is the phase of DM, $\epsilon$ is the interaction strength between DM and sensors, and $\sigma^{A}_i$ is the Pauli-$A$ matrix applied to the $i$th qubit ($A=X,Y,Z$).
The interaction strength $\epsilon$ is determined by the coupling constant between sensors and DM and the DM field amplitude and we would like to estimate it from the measurement of the qubits.

Let us move to the interaction picture with respect to $H_0$.
In the following, we focus on the resonant condition where the qubit frequency and the mass of DM are close together, $\omega \sim m$, and neglect the fast oscillating term, assuming $t (\omega + m) \gg 1$. 
Then, the Hamiltonian in the interaction picture is
\begin{equation}
    H_I=
    \epsilon  \sum_i \left[ \sigma^X_i \cos( \Delta \omega t+\varphi) + \sigma^Y_i \sin( \Delta \omega t+\varphi) \right]
    \label{eq:HI}
\end{equation}
where $\Delta \omega \equiv \omega-m$ is the detuning between the qubit frequency $\omega$ and the DM signal frequency (derived from its mass).

With noise effects, the state of the system is not pure anymore, and we need to describe it by the density matrix $\rho(t)$.
The time evolution of the density matrix with noise effects is described by the Lindblad equation as:
\begin{equation}
    \frac{d \rho }{dt}= -i[H_I, \rho]+\sum_i D_i[\rho], \label{eq:general_Lindblad}
\end{equation}
where $D_i[\rho]$ are Lindblad superoperators describing the noise effects. See Appendix~\ref{sec:appendix_Lindblad} for a brief review of the Lindblad equation. 
For the noise effects, we consider the excitation, amplitude damping, and dephasing effects of qubits.
They are described by the following Lindblad superoperators, $D_0$, $D_1$, and $D_2$, respectively:
\begin{gather}
   D_0[\rho] \equiv \Gamma_0 \sum_i \left(  \sigma^-_i \rho \sigma^+_i-  \frac{1}{2}\{\sigma^+_i\sigma^-_i,\rho \} \right), \label{eq:ex}\\
    D_1[\rho] \equiv \Gamma_1 \sum_i \left(  \sigma_i^+ \rho \sigma^-_i- \frac{1}{2}\{  \sigma_i^- \sigma_i^+,\rho \}\right),  \label{eq:ampdamp}\ \\
    D_2[\rho] \equiv \frac{\Gamma_2}{2} \sum_i \left(  \sigma^Z_i \rho \sigma^Z_i- \rho \right) ,\label{eq:phasedamp}
\end{gather}
where $\Gamma_{0,1,2}$ are the rates of each effect, and $\sigma_i^{\pm} \equiv \frac{1}{2}( \sigma_i^X\pm i \sigma_i^Y)$ so that $\sigma_i^-=\ketbra{1}{0}_i$ and $\sigma_i^+=\ketbra{0}{1}_i$. Here, for simplicity, we assume that the rates are the same for all qubits.
The excitation noise, $D_0$, given by Eq.~\eqref{eq:ex} describes the background excitation of qubits from the ground state to the excited state, which mimics the DM signal, and the deexcitation noise, $D_1$, given by Eq.~\eqref{eq:ampdamp} describes the amplitude damping of qubits from the excited state to the ground state. The phase damping noise, $D_2$, together with the excitation and amplitude damping, describes the decoherence effects of qubits.
In the following, we consider the situation where the excitation noise is much smaller than the amplitude damping and dephasing noise, $\Gamma_0 \ll \Gamma_1, \Gamma_2$, which is the case for most qubit sensors at present~\cite{Jin_2015,ibm_quantum,Piskor:2025xla,Figueroa-Romero:2024dcl}.

In addition to the noise effects of the sensors, we also need to consider the coherence time scale of the DM signal itself.
Typically, the DM field coherently oscillates within the time scale determined by~\cite{Cheong:2024ose}
\begin{equation}
    \tau_{\rm DM}\sim \frac{1}{mv^2},
\end{equation}
within which the phase $\varphi$ of DM can be regarded as a constant (but random) value, where $v\sim 10^{-3}$ is the DM velocity. 
In this paper, we focus on the situation where the sensors are noisy and we assume the decoherence time scale of sensors is much shorter than the DM coherence time scale, ignoring the effect of the DM decoherence.

Next, we introduce the protocol to measure the DM signal and define the sensitivity. We consider initializing all qubit sensors in the ground state, i.e., 
\begin{align}
    \rho(0)= \ketbra{i}{i}, ~~~\mbox{with}~~~\ket{i}=\ket{0}^{\otimes L},
    \label{eq:initial}
\end{align}
and measure its transition to a certain final state $\ket{f}$. The probability $p$ of the measurement is given by
\begin{equation}
    p=\mathrm{Tr}[P_f\rho(t)],
\end{equation}
where the projection operator can be written as
\begin{equation}
    P_f\equiv\ketbra{f}{f}.
\end{equation}

From the measurement of the projection operator $P_f$, we can derive the information of the signal parameter $\epsilon$. 
In order to quantify the sensitivity of the measurement, we use the uncertainty of the parameter estimation, $\delta \epsilon$, in quantum metrology, which is approximated as
\begin{equation}
    \label{eq:stddev_epsilon}
    \delta \epsilon 
    =
    \frac{\delta p}{|d p/d\epsilon|}= \frac{1}{\sqrt{N}}\frac{\sqrt{p-p^2}}{|d p/d\epsilon|},
\end{equation}
where, in the last equation, we used that the standard deviation of the observable $p$ is given by
\begin{equation}
    \delta p = \sqrt{\frac{p(1-p)}{N}},
\end{equation}
where $N$ is the number of measurements.
Note that the uncertainty $\delta \epsilon$ is the standard deviation of the estimator of the parameter $\epsilon$; therefore, the smaller $\delta \epsilon$ is, the better sensitivity we have. In the context of high-energy physics, the signal-to-noise ratio is often used to quantify the sensitivity, which corresponds to the likelihood ratio test assuming Gaussian statistics.
In general, we should consider the standard deviation of the estimator of the parameter $\epsilon$, $\delta \epsilon$, to quantify the sensitivity of the measurement. 
We review the derivation of Eq.\,\eqref{eq:stddev_epsilon} and
the relation between $\delta \epsilon$ and the signal-to-noise ratio in Appendix~\ref{sec:appendix_parameter_estimation}.

\section{Dark matter detection with separate measurements}
In this section, we solve the Lindblad equation, Eq.~\eqref{eq:general_Lindblad}, to follow the time evolution of the density matrix of qubit sensors with the DM effect and noise effects. 
Then, we focus on the case when each qubit is measured separately to observe the effect of DM.

In the case of our interest, the entanglement between qubits does not exist; the Lindblad equation \eqref{eq:general_Lindblad} as well as the initial condition given in Eq.~\eqref{eq:initial} guarantees that the total density matrix can be expressed by the tensor product of each qubit's density matrix $\rho_i$ as
\begin{align}
    \rho = \bigotimes_{i=1}^L \rho_i.
\end{align}
The density matrix of an individual qubit evolves independently to others, so we can focus on the time evolution of a single qubit.

To solve the Lindblad equation, we expand the density matrix as
\begin{equation}
    \rho_i=
        c_{00}(t) \ketbra{0} + e^{-i\varphi }c_{01}(t) \ketbra{0}{1}
        +e^{i\varphi }c_{10}(t)  \ketbra{1}{0}+ c_{11}(t) \ketbra{1},\label{eq:density matrix}
\end{equation}
where $c_{00}=1-c_{11}$ and $c_{01}=c_{10}^*$. Here, we extract the phase factor $e^{\pm i\varphi}$, which originates from the unknown phase of DM, from the off-diagonal components of the density matrix, which simplifies the equations of motion.
Taking into account the background excitation and other damping effects given by Eqs.~\eqref{eq:ex}, \eqref{eq:ampdamp} and \eqref{eq:phasedamp}, respectively, the equation of motion, Eq.~\eqref{eq:general_Lindblad}, for each component of the density matrix $\rho_i$ reads as
\begin{equation}
    \begin{pmatrix}
    \dot{c}_{00}\\
    \dot{c}_{01}\\
    \dot{c}_{10}\\
    \dot{c}_{11}
    \end{pmatrix}
    =
    \begin{pmatrix}
    -\Gamma_0&+ i \epsilon e^{-i\Delta \omega t}   &-i\epsilon e^{i\Delta \omega t}   &+\Gamma_1\\
    + i\epsilon   e^{i\Delta \omega t} &-\frac{\Gamma_0+\Gamma_1}{2}- \Gamma_2 &0&-i\epsilon e^{i\Delta \omega t}  \\
     -i\epsilon e^{-i\Delta \omega t}  &0&-\frac{\Gamma_0+\Gamma_1}{2}- \Gamma_2 &+ i\epsilon e^{-i\Delta \omega t}  \\
    +\Gamma_0& -i\epsilon e^{-i\Delta \omega t}  & + i\epsilon e^{i\Delta \omega t} &-\Gamma_1
    \end{pmatrix}
    \begin{pmatrix}
    c_{00}\\
    c_{01}\\
    c_{10}\\
    c_{11}
    \end{pmatrix}. \label{eq: EOMc}
\end{equation}
To solve the equation of motion, we expand the solution perturbatively in terms of the small parameter $\epsilon t \ll 1$, whereas we do not assume that the decoherence rates $\Gamma_{0,1,2} t$ are small. The equation is solved by the perturbative expansion as shown in Appendix~\ref{sec:appendix_perturbation}.
Since the initial state is $\ket{0}$, we obtain the solution at later time $t$ as
\begin{align}
     c_{01}(t) &=  \frac{i\epsilon}{\gamma_1} e^{-\gamma_2 t} \left[2\Gamma_0 \frac{1 - e^{(-\gamma_1+\gamma_{2+}) t}}{\gamma_1-\gamma_{2+}} 
    + (\Gamma_1-\Gamma_0) \frac{e^{\gamma_{2+} t}-1}{\gamma_{2+}}  \right] + \mathcal{O}(\epsilon^3), \label{eq:c01} \\ 
    c_{10}(t) &=  \qty(c_{01}(t))^\star, \label{eq:c10}
\end{align} 
and
\begin{align}
    c_{11}(t) =& (1-e^{-\gamma_1 t})
    \frac{\Gamma_0}{\gamma_1}\nonumber \\
    &+\sum_{s=\pm} \frac{\epsilon^2}{\gamma_1\gamma_{2s}}\Bigg\{(1-e^{-\gamma_1 t})  \frac{\Gamma_1-\Gamma_0}{\gamma_1} \nonumber \\
    & \quad \quad + \frac{\Gamma_0}{\gamma_1-\gamma_{2s}} \qty[
   (e^{-\gamma_{2s} t} - e^{-\gamma_1 t}) \qty(\frac{\gamma_1+\gamma_{2s}}{\gamma_1-\gamma_{2s}} -\frac{\Gamma_1}{\Gamma_0}) -2  \gamma_{2s}t e^{-\gamma_1 t}
   ]\Bigg\} + \mathcal{O}(\epsilon^4)\label{eq:c11} 
\end{align}
where we define
\begin{gather}
    \gamma_1\equiv \Gamma_0+\Gamma_1,\\
    \gamma_{2s} \equiv \gamma_2 + s i \Delta \omega, \\
    \gamma_2\equiv\Gamma_2+\frac{(\Gamma_0+\Gamma_1)}{2}.
\end{gather}
Note that, for $\Delta \omega=0$, $\gamma_1$ and $\gamma_{2}$ correspond to the longitudinal and transverse relaxation rate, respectively; the longitudinal relaxation time $T_1$ and the transverse relaxation time $T_2$ are given by $T_1=1/\gamma_1$ and  $T_2=1/\gamma_2$. 
Here, the density matrix is expanded in terms of $\epsilon$. Because the density matrix reduces to the asymptotic one as $t\gg\gamma_1^{-1}$ (assuming $\gamma_1\lesssim\gamma_2$), such an expansion is valid when $\epsilon\ll\gamma_1$.

Now, let us consider the case of the separate measurement,i.e., the case that each qubit is measured separately.
The phase of the DM, $\varphi$, is unknown and can be treated as a random variable uniformly distributed in $[0,2\pi)$. Therefore, to measure the DM signal from each qubit, Eq.\,\eqref{eq:density matrix}, we need to extract $c_{11}$; i.e., we perform the projection measurement to the excited state, $P_f=\ketbra{1}{1}$. 
The probability $p_1$ that a qubit is excited is then given by
\begin{equation}
    p_1(t)=c_{11}(t)\equiv p_{1,\rm BG}+p_{1,\rm sig},
\end{equation}
where we separated $p_1$ into the background part $p_{1,\rm BG}$ 
and the signal part $p_{1,\rm sig}$ for convenience.
The background part is independent of $\epsilon$ and comes from the excitation noise, $\Gamma_0$, while the signal part is proportional to $\epsilon$.
In particular,
\begin{align}
    p_{1,\rm BG}&=(1-e^{-\gamma_1 t}) \frac{\Gamma_0}{\gamma_1}, \label{eq:p1BG}\\
    p_{1,\rm sig} &\simeq \sum_{s=\pm} \frac{\epsilon^2}{\gamma_1\gamma_{2s}}\Bigg\{(1-e^{-\gamma_1 t})  \frac{\Gamma_1-\Gamma_0}{\gamma_1} \nonumber  \\
    & \quad \quad + \frac{\Gamma_0}{\gamma_1-\gamma_{2s}} \qty[
   (e^{-\gamma_{2s} t} - e^{-\gamma_1 t}) \qty(\frac{\gamma_1+\gamma_{2s}}{\gamma_1-\gamma_{2s}} -\frac{\Gamma_1}{\Gamma_0}) -2  \gamma_{2s}t e^{-\gamma_1 t}
   ]\Bigg\}.
\end{align}

Before discussing the parameter estimation, let us comment on the behavior of the signal at several relevant time scales, and the bandwidth of the measurement; i.e., the range of the detuning $\Delta \omega$, $\Delta \omega_{\rm BW}$, within which we may receive the signal. If we take the limit $\gamma_1 \gg t^{-1}$ and $\gamma_2 \gg t^{-1}$, where the noise effects are strong enough to saturate the system within time $t$, the signal part $p_{1,\rm sig}$ is approximated as
\begin{align}
     p_{1,\rm sig} &\simeq \frac{2 \epsilon^2 (\Gamma_1-\Gamma_0) \gamma_2}{\gamma_1^2 (\gamma_2^2+\Delta \omega^2)}.
        \label{eq:sig_strong_noise}
\end{align}
Next, in the case where $\gamma_2\gg\gamma_1$ and $1/\gamma_2 \ll t\ll 1/\gamma_{1}$, $p_{1,\rm sig}$ is approximated as
\begin{equation}
    p_{1,\rm sig}\simeq\frac{2\epsilon^2  \gamma_2}{\gamma_2^2+\Delta\omega^2}t. \label{eq:middle}
\end{equation}
On the other hand, if we take the limit $\gamma_1 \ll t^{-1}$ and $\gamma_2 \ll t^{-1}$, 
where the noise effects are negligible within time $t$, the signal part is approximated as
\begin{align}
    p_{1,\rm sig} &\simeq \epsilon^2 t^2 \text{sinc}^2\qty(\frac{t\Delta \omega}{2}),
    \label{eq:sig_weak_noise}   
\end{align}
where $\text{sinc}(x)\equiv \sin x/x$.
For the bandwidth, one can see that, when $t \gg 1/\gamma_2$, the effect of the detuning $\Delta \omega$ is ignorable if $\Delta \omega \lesssim \gamma_2$, while for the opposite limit when $t\ll 1/\gamma_2$, the effect of the detuning is ignorable if $\Delta \omega \lesssim 1/t$.
Therefore, the bandwidth of the measurement is well approximated by
\begin{equation}
    \Delta \omega_{\rm BW} \sim \max \left( \frac{1}{t},\gamma_2\right). \label{eq:BW}
\end{equation}
In other words, for a given bandwidth $\Delta \omega_{\rm BW}$ and the qubit frequency $\omega$, the measurement probe the DM with mass in range $ \omega-\Delta \omega_{\rm BW}  \lesssim m_{\rm DM} \lesssim \omega+\Delta \omega_{\rm BW}$.
In the rest of this section, we assume $\Delta \omega \lesssim \Delta \omega_{\rm BW}$ and ignore the detuning $\Delta \omega$.

Next, we derive the uncertainty of the $\epsilon$ parameter estimation from the measurement of $L$ in parallel.
We then determine the optimal measurement time $t$ to minimize the uncertainty.
With the total observation time $T$ fixed, the total repetitions of the measurement of all qubits in this case is equal to $N^{(\rm sep)}= L T/t$ for a given measurement time $t$.
Then, we obtain the standard deviation of the estimator of $\epsilon$ as
\begin{equation}
    \delta \epsilon^{(\rm sep)}= \frac{1}{\sqrt{N^{\rm(sep)}}}\frac{\sqrt{p_1(1-p_1)}}{\qty|dp_1/d\epsilon|}.
\end{equation}
See Appendix~\ref{sec:appendix_parameter_estimation} for the detail.
Assuming the signal is smaller than the background, $\epsilon^2 \ll \Gamma_0 \gamma_2$, the uncertainty can be written as
\begin{equation}
    \delta \epsilon^{\rm (sep)}
    \simeq \frac{1}{2}\frac{1}{\sqrt{L}}
    \sqrt{\frac{\Gamma_0}{T}}
    \frac{\gamma_2}{2 \epsilon} \left[1- 
    \frac{\gamma_2e^{-\gamma_1 t} -\gamma_1e^{-\gamma_2 t}}{\gamma_2-\gamma_1}
    \right]^{-1}  \sqrt{\gamma_1 t(1-e^{-\gamma_1 t}) },
\end{equation}
where we also used $\Gamma_0 \ll \gamma_1$.
Let us discuss the optimal time $t$ to minimize the uncertainty.
For $\gamma_1\sim \gamma_2$, the uncertainty is minimized at $t\sim 1/\gamma_2$.
With optimized time $t$ chosen for each measurement, one obtains the uncertainty as
\begin{equation}
    \delta \epsilon^{\rm (sep)}
    \sim
    \frac{1}{\sqrt{L}}
    \sqrt{\frac{\Gamma_0}{T}}
    \frac{\gamma_2}{ \epsilon} \label{eq:desep}
\end{equation}
up to an $O(1)$ prefactor.
On the other hand, when $\gamma_2 \gg \gamma_1$, the uncertainty is minimized approximately equally for any measurement time within the interval $1/\gamma_2 \lesssim t \lesssim 1/\gamma_1$. There, the uncertainty is analytically equal to Eq.~\eqref{eq:desep} up to an $O(1)$ prefactor, exhibiting a plateau independent of time within $1/\gamma_2 \lesssim t \lesssim 1/\gamma_1$. That is because both the background excitation and the signal grow linearly with time as discussed earlier, Eqs.~\eqref{eq:p1BG} and \eqref{eq:middle}, and the shorter (longer) measurement times $t$ within the interval, compensated precisely by a larger (smaller) number of repetitions, yield the same uncertainty.

\section{Dark matter detection by projecting qubit sensors to $W$ state}

In this section, instead of separate measurements, we consider the situation where we perform a collective measurement of all qubits, projecting them into an entangled state (particularly, the $W$ state, which we shall explain shortly).
We then show that the background excitation can be significantly suppressed.
This is because while the signal is applied coherently between sensors, the background noises occurring on the sensors are not correlated.
In other words, if we carefully choose a measurement operator projecting the state of sensors into a ``signal'' subspace, where the sensor state evolves under the effect of DM, the noise effects can be significantly suppressed; while DM affects qubits and contributes directly to the collective excitation, the noise instead contributes mainly to the change of qubits to other subspaces.

We specifically consider the projection operator projecting the state to the so-called $W$ state:
\begin{equation}
    P_W=\ketbra{W}{W},
\end{equation}
where 
\begin{equation}
    \ket{W}=\frac{1}{\sqrt{L}}(\ket{0\cdots001}+\ket{0\cdots010}+\ket{0\cdots100}+\cdots+\ket{10\cdots0}),
\end{equation}
which is the superposition of the qubit states with only one excitation.
The probability of the projection is given by
\begin{align}
    p_W \equiv \mathrm{Tr}[\rho(t)P_W] &= \expval{\rho(t)}{W}. \label{eq:pw}
\end{align}
The difference between this protocol and the separate measurements in the previous section lies merely in the measurement, where, in this case, we perform the measurement projecting the state of sensors into the entangled state.
The state of sensors at time $t$ before the measurement is as the previous case and simply the tensor product between those sensors:
\begin{equation}
    \rho(t)= \rho_1(t) \otimes \rho_2(t) \otimes \cdots \otimes \rho_L(t)
\end{equation}
where each qubit density matrix $\rho_i(t)$ is given by Eq.~\eqref{eq:density matrix}. 
Since the projection of $W$ state only includes the exact one excitation, let us focus on those relevant terms of the density matrix:
\begin{equation}
    \rho(t) 
    \supset 
    c_{00}^{L-1}c_{11}
    \sum^L_{i=1} \sigma^X_i \ketbra{0\cdots0} \sigma^X_i
    +
    c_{00}^{L-2}c_{01}c_{10}
    \sum^L_{i \neq j} \sigma^X_i \ketbra{0\cdots0} \sigma^X_j.
\end{equation}
Importantly, not only the $c_{11}$ term contribute to the projection probability, the $c_{01},c_{10}$ also contribute to it.
Then, the probability of the projection into the $W$ state can be directly calculated using Eq.~\eqref{eq:pw}. 
 It can be separated as a sum of the background and signal parts, $p_{ W}=p_{ W,\rm BG}+p_{ W,\rm sig}$ with 
\begin{align}
    p_{W, \rm BG} &= (1-p_{1,\rm BG})^{L-1} p_{1,\rm BG}, \label{eq:pWbg}\\
    p_{W,\rm sig}& = (1-p_{1,\rm BG})^{L-1}p_{1,\rm sig}+(1-p_{1,\rm BG})^{L-2}(L-1)|c_{01}|^2 \nonumber\\
    &\quad\quad -(1-p_{1,\rm BG})^{L-2}(L-1)p_{1,\rm BG}p_{1,\rm sig} + O(\epsilon^3), \label{eq:pWsig}
\end{align}
where we keep only the leading order terms with respect to $\epsilon$ for the signal part. 
To treat $\epsilon$ perturbatively in this formula, we need to assume $p_{1,\rm BG} \gtrsim L |c_{01}|^2$ in addition to the condition in the previous section, $\epsilon t \lesssim 1$.
Hereafter, we always assume this condition. Given that, let us comment on the structure of the above results.

First, the $W$ state includes only one excited qubit. Thus, the projection probability onto the $W$ state becomes suppressed when $p_{1,\rm BG}$ is sizable because two or more qubits may be excited by the background excitation process.
One can see this from Eq.\,\eqref{eq:pWsig}. The first two terms are suppressed by $\mathcal{O}\qty((1-p_{1,\rm BG})^{L})$, which is the probability that each qubit is not excited by the background noise. The last term is negative, which also reflects the suppression of the projection probability due to the background excitation.
Note that the factor $\mathcal{O}\qty((1-p_{1,\rm BG})^{L})$ becomes important when 
$p_{1,\rm BG} \gtrsim 1/L$.
For $L\Gamma_0/\gamma_1 \gtrsim 1$, this factor becomes important at time scale $t \sim 1/L\Gamma_0$; therefore, the measurement time should be shorter than this to avoid the suppression of the signal from the background excitation. On the other hand, for $L\Gamma_0/\gamma_1 \lesssim 1$, this factor is negligible for an arbitrary $t$.

Second, assuming that $L\Gamma_0/\gamma_1 \lesssim 1$ and the factor $\mathcal{O}\qty((1-p_{1,\rm BG})^{L})$ is close to unity, let us discuss the behavior of the background and signal parts.
The background part $p_{W,\rm BG}$ is approximately equal to $p_{1,\rm BG}$, which is the background excitation of a single qubit. However, the signal part $p_{W,\rm sig}$ includes the term proportional to $|c_{01}|^2$, which increases with $L$.
Therefore, in this way, by projecting the state of the sensors to the $W$ state, we can suppress the background excitation significantly and enhance the sensitivity for DM detection.
We note that even though the negative term in Eq.~\eqref{eq:pWsig} also depends on $L$, for the relevant time scales, it is much smaller than the term proportional to $|c_{01}|^2$ due to a small factor $p_{1,\rm BG}$.

Third, let us comment on the bandwidth of the measurement for  $\mathcal{O}\qty((1-p_{1,\rm BG})^{L}) \sim 1$. In this case, as discussed above, the behavior of $p_{W,\rm sig}$ is dominated by the term proportional to $|c_{01}|^2$. Similar to the separate measurement case, for the larger noise effects, $\gamma_1 \gg t^{-1}$ and $\gamma_2 \gg t^{-1}$,
\begin{align}
    |c_{01}|^2 &\simeq \frac{\epsilon^2(\Gamma_0 - \Gamma_1)^2}{\gamma_1^2 (\gamma_2^2+\Delta \omega^2)},
\end{align}
while for the smaller noise effects, $\gamma_1 \ll t^{-1}$ and $\gamma_2 \ll t^{-1}$,
\begin{align}
    |c_{01}|^2 &\simeq \epsilon^2 t^2 \text{sinc}^2\qty(\frac{t\Delta \omega}{2}).
\end{align}
Therefore, the bandwidth of the measurement is the same as the separate measurement case and is given by Eq.~\eqref{eq:BW}, $\Delta \omega_{\rm BW} \sim \max \left( \frac{1}{t},\gamma_2\right)$.
We will comment on the frequency scanability for practical DM search at the end of this section.
In the rest of this section, we again assume $\Delta \omega \lesssim \Delta \omega_{\rm BW}$ and ignore the detuning $\Delta \omega$.

Next, we analyze the uncertainty of DM parameter estimation in this case, where one uses $L$ qubits and projects it to the $W$ state. For the fixed total observation time $T$ as in the separate measurement, with each measurement costing time $t$, we can repeat the measurement for $N^{(W)}=T/t$ times.
Then, we obtain the uncertainty in this case as
\begin{equation}
    \delta \epsilon^{(W)}= \frac{1}{\sqrt{N^{(W)}}} \frac{\sqrt{p_W(1-p_W)}}{|dp_W/d\epsilon|}.
\end{equation}
This is to be compared with the uncertainty of the separate measurement case, $\delta \epsilon^{(\rm sep)}$, given in the previous section.

Let us first discuss the analytic result before showing the numerical result.
To understand the behavior of the uncertainty as a function of $L$, let us consider the limit when the number of qubits is much larger than one, $L\gg 1 $.
We also assume that the background excitation is larger than the signal, $p_{W,\rm BG} \gg p_{W,\rm sig}$, and the background excitation is small enough, $p_{W,\rm BG} \ll 1$.
The uncertainty can be approximated by
\begin{equation}
    \delta \epsilon^{(W)} \simeq \frac{1}{2}\frac{1}{L} \sqrt{\frac{\Gamma_0}{T}} \frac{\gamma_2}{\epsilon} \frac{\gamma_2/\gamma_1}{(1-e^{-\gamma_2t})^2} \sqrt{\gamma_1t(1-e^{-\gamma_1t})} \left( 1-(1-e^{-\gamma_1t}) \frac{\Gamma_0}{\gamma_1}\right)^{-L/2}.
\end{equation}
Now, let us discuss the measurement time $t$ that minimizes the uncertainty $\delta \epsilon^{(W)}$.
The behavior of the uncertainty is determined by the competition between the accumulation of the signal and the suppression of the signal from the background excitation. 
As discussed earlier, the former time scale is $\sim1/\gamma_2$ while the latter depends on whether $L\Gamma_0/\gamma_1$ is smaller or larger than 1.
First, in the case where $L\Gamma_0/\gamma_1 \ll 1$, the suppression of the signal from the background excitations is negligible, so the uncertainty is simply minimized when the signal is accumulated fully, which is at the time $t\sim 1/\gamma_2$.
Next, we turn to the case when $L\Gamma_0/\gamma_1 \gg 1$, for which the suppression of signal becomes important when $t\sim1/L\Gamma_0$. Then, depending on whether the signal accumulates before or after the suppression effect due to the background excitations, the time $t$ to minimize the uncertainty is determined, which is given by $t\sim \min\{1/\gamma_2,1/L\Gamma_0\}$.
Assuming the optimized time is chosen and taking into account that $\gamma_2\gtrsim \gamma_1$, we have the uncertainty following
\begin{equation}
    \delta \epsilon^{(W)}\sim
    \begin{dcases}
         \frac{1}{L} \sqrt{\frac{\Gamma_0}{T}} \frac{\gamma_2}{\epsilon}
          & \text{for} \quad L\Gamma_0/\gamma_2 \lesssim 1\\
          \sqrt{\frac{\Gamma_0}{T}} \frac{\Gamma_0}{ \epsilon}
         & \text{for} \quad  L\Gamma_0/\gamma_2 \gtrsim 1
    \end{dcases} \label{eq:depW}
\end{equation}
up to a prefactor of $\sim O(1)$.
Comparing it to the result of the separate measurements, we obtain
\begin{equation}
    \frac{\delta \epsilon^{(W)}}{\delta \epsilon^{(\rm sep)}} \sim 
    \begin{dcases}
        \frac{1}{\sqrt{L}} & \text{for} \quad L\Gamma_0/\gamma_2 \lesssim 1\\
        \sqrt{L}\frac{\Gamma_0 }{\gamma_2} & \text{for} \quad  L\Gamma_0/\gamma_2 \gtrsim 1
    \end{dcases}
\end{equation}
up to a prefactor of $\sim O(1)$.
Compared to the separate measurement, the performance when using $W$ state grows with better scaling with respect to the number $L$ of qubits to some point around $L\Gamma_0/\gamma_2\sim1$ and then goes down after that.
This is because, for $L\Gamma_0/\gamma_2 \lesssim 1$, the background excitation is not so significant and the signal accumulation dominates the behavior of the uncertainty. On the other hand, for $L\Gamma_0/\gamma_2 \gtrsim 1$, the background excitation becomes significant and suppresses the signal, leading to the increase of the uncertainty. Note that for $L\Gamma_0/\gamma_2 \gtrsim 1$, we may separate $L$ qubits into $N_L \sim L \Gamma_0/\gamma_2$ groups, each of which includes $L' \sim \gamma_2/\Gamma_0$ qubits, and perform the $W$ state projection for each group separately. Then, the uncertainty is reduced by a factor of $\sqrt{L'}$ compared to the separate measurement, and we may avoid the increase of the uncertainty for $L\Gamma_0/\gamma_2 \gtrsim 1$.
Assuming $L\Gamma_0/\gamma_2 \lesssim 1$ is satisfied, our perturbative treatment is valid up to $L \lesssim \Gamma_0\gamma_2 / \epsilon^2$. 
(Note that for DM detection, the coupling $\epsilon$ is generically so small that $\Gamma_0\gamma_2 / \epsilon^2\gg 1$. Thus, the perturbative treatment can be applicable for $L\gg 1$.)
At this point, we can show that our protocol using the $W$ state is \emph{optimal} in the sense that it saturates the quantum Cram\'er-Rao bound~\cite{Paris:2008zgg}. See Appendix~\ref{app:QCRB} for the details.

\begin{figure}[t]
    \centering
    \includegraphics[width=0.7 \textwidth]{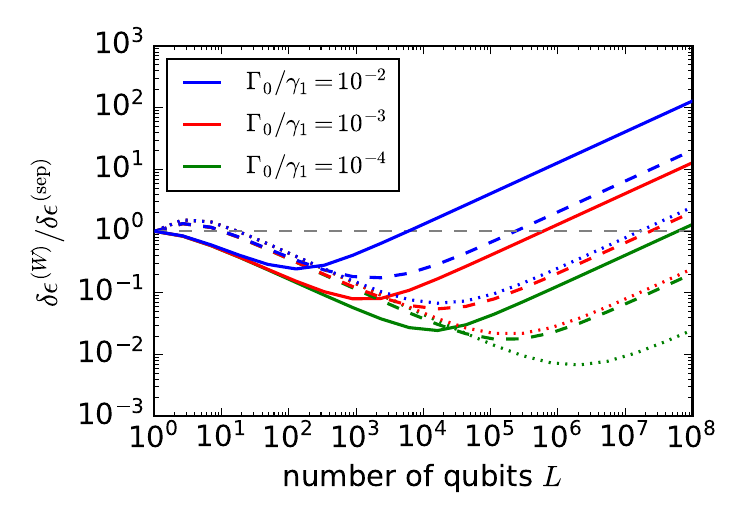}
    \caption{The ratio $\delta \epsilon^{ (W)}/\delta \epsilon^{\rm (sep)}$ of the uncertainties of measuring the DM signal between the case using $W$ state and the separate measurement. The solid, dashed, and dotted lines correspond to $\gamma_2/\gamma_1 \simeq 1,10$ and $100$, respectively. }
    \label{fig:gain}
\end{figure}

In Fig.~\ref{fig:gain}, we show the ratio of the uncertainties $\delta \epsilon^{(W)}/\delta \epsilon^{(\rm sep)}$ as a function of $L$ for several choices of $\gamma_2/\gamma_1$.
(As a numerical reference value, for qubit sensors such as transmon superconducting qubits, we expected background excitation probability $10^{-3}$~\cite{ibm_quantum,Piskor:2025xla,Figueroa-Romero:2024dcl} which corresponds to $\Gamma_0/\gamma_1\sim 10^{-3}$. On the other hand, the ratio $\gamma_2/\gamma_1$ is determined by the ratio of the relaxation times $T_1$ and $T_2$.)
Here, we numerically optimize the measurement time $t$ for each case to minimize the uncertainty. Also, we take all terms in Eq.~\eqref{eq:pWsig} into account, not only the leading order terms with respect to $L$.
One can see that the behavior is consistent with the above analytical argument; the optimal choice of number of qubits is around $L\sim \gamma_2/\Gamma_0$ with the suppression factor as $\sim\sqrt{\Gamma_0/\gamma_2}$, while the scaling with respect to number of qubits $L$ is also consistent with the analytical argument.

\begin{figure}[t]
    \centering
    \includegraphics[width=0.7 \textwidth]{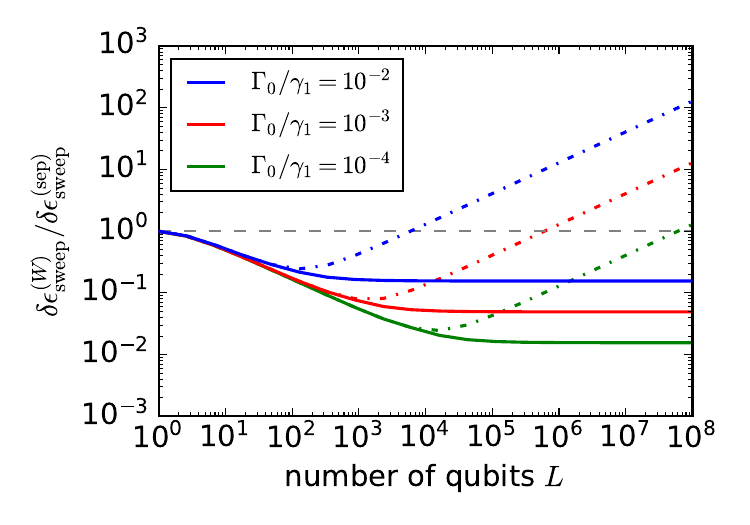}
    \caption{The ratio $\delta \epsilon_{\rm sweep}^{ (W)}/\delta \epsilon_{\rm sweep}^{\rm (sep)}$ of the uncertainties of measuring the DM signal with unknown frequency between the case using $W$ state and the separate measurement with $\gamma_2/\gamma_1=1$. The dash-dotted lines are the ratio for the case of a signal with a known frequency provided as reference lines. The uncertainty of $W$ state case is supported by the wider bandwidth for $ L  \gtrsim \gamma_2/\Gamma_0$.}
    \label{fig:gain_sweep}
\end{figure}

Before leaving this section, we add a discussion on the situation when the frequency of the signal is unknown and we need to scan a certain frequency range of interest. This situation is directly applied to the practical execution of the protocol for the DM search, where one does not know the DM signal frequency exactly.
We show that in this case, the benefit of projecting the sensors to the $W$ state compared to the separate measurement remains, and even better for the large $L$ limit, owing to the wider bandwidth of the measurement.
We emphasize that the protocol offers the crucial advantage of non-collective background suppression, while supporting frequency scanning in the similar manner as standard resonance experiments, e.g., superconducting qubit DM searches~\cite{Chen:2022quj}.

Suppose we need to scan a frequency range $F$ to search for a signal with an unknown frequency.
Here, we assume the frequency-tunable quantum sensor and simply adopt the protocol of dividing the total frequency range $F$ to be searched to a number of bins of equal bin-width and arranging the observation time for each bin equally.
The bin-width $b$ is chosen by the bandwidth receiving signal, $b=\Delta \omega_{\rm BW}$, where the bandwidth $\Delta \omega_{\rm BW}$ is given by Eq.~\eqref{eq:BW}.
In this case, where one distributes the total observation time $T$ equally among all $F/b$ number of bins, the repetition times of measurements at each bin are determined by
\begin{equation}
N^{(\rm sep)} = L \frac{T}{ t} \frac{b}{F} , \quad N^{(W)} = \frac{T}{ t} \frac{b}{F}
\end{equation}
for the case of separate measurement and $W$ state projection, respectively.
We again numerically optimize the measurement time $t$ for each case and compare their uncertainties $\delta \epsilon$ of the parameter estimation. 
(The optimized times differ from those of the previous case when we know a signal frequency by only a small prefactor.)
The comparison between the uncertainty $\delta \epsilon$ using the $W$ state and separate measurement is plotted in Fig.~\ref{fig:gain_sweep}.  
One can see that for $L \lesssim \gamma_2/\Gamma_0$ the behavior is the same as the case when one knows the signal frequency. On the other hand, for large $L$ satisfying $L\gtrsim \gamma_2/\Gamma_0$, the scaling with respect to $L$ is improved from the case where the frequency is known, since in this case the effect of bandwidth becomes important.
 This is when the uncertainty is minimized with time determined by $t\sim 1/L \Gamma_0$, determined by the time scale where the background excitations of all sensors become important. 
 In this case, even though the measurement time is shorter by the factor $1/L$, the sensitivity of the $W$ state case is compensated by the wider bandwidth, $\Delta \omega_{\rm BW} \sim 1/t \propto L$. In other words, there, we have a smaller number of total bins, giving the longer observation time that can be spent at each bin, effectively compensating for the shorter measurement time $t$,  given a fixed total time $T$ and fixed range $F$. 
 A similar benefit relying on the wide bandwidth to search for the signal with unknown frequency, is also discussed in Ref.~\cite{Sichanugrist:2024wfk} on using the highly entangled state called the GHZ state~\cite{Greenberger:1989tfe}.

\section{Discussion and Conclusion}
In this work, we proposed a protocol to enhance the sensitivity for DM direct detection experiments using qubit sensors by projecting the state of sensors into the collective excitation state ($W$ state).
We showed that our protocol can significantly suppress the effect of the background excitation by a factor equal to the number of qubit sensors and enhance the sensitivity for DM direct detection experiments using qubit sensors.
Our protocol does not require entanglement of sensors during the signal accumulation, hence avoiding the difficulty of maintaining the long coherence time of the entangled sensors.
We take into account the noise effects, including background excitation, amplitude damping, and dephasing, optimize the measurement time of the protocol and demonstrate the advantage of our protocol compared to the separate measurement of each qubit sensor, as well as its limitations.

In particular, quantitatively, for a setup with the typical background excitation probability of 0.1\% with a number of qubits around $10^2-10^3$, one can reduce the uncertainty for the DM parameter estimation from the measurement by a factor of $10^{-1}\sim 10^{-2}$ compared to the separate measurement.  
For the instance of using superconducting transmon qubits for probing hidden photon DM~\cite{Chen:2022quj}, scanning DM mass from $4- 40 \ \rm \mu eV$ for a 1 year plan, we expect the sensitivity down to $\epsilon \simeq 10^{-14}$ to be achievable, given $ O(100)$ qubits and high-fidelity $W$ state projection.

One technical challenge lies in the performance of the projection to the $W$ state. 
The imperfection of the $W$ state projection becomes an additional source of noises. 
The background excitation suppression from the protocol remains effective, giving the same suppression of the uncertainty of the parameter estimation compared to the separate measurement by the scaling of $O(1/\sqrt{L})$; however, the low fidelity of the $W$ state readout would suppress the observed signal directly. 
The optimized gate sequence~\cite{Plesch:2011vwn,Zhang:2022pue,Wagner:2024ckk} 
might help mitigate these effects.

In addition, the $W$-state projection typically requires gate operations up to $O(L)$ gates/steps, so its operation time could be non-negligible. 
(For the qubit preparation in the ground state of the protocol,  even if we increase the number of qubits, it can be done in parallel and without entanglement, so the qubit preparation time can be negligible compared to the signal accumulation time, as estimated in Ref.~\cite{Chen:2022quj}.)
For superconducting qubits, the gate operation time is of $O(10-100) \ \mathrm{ns}$~\cite{ibm_quantum,Piskor:2025xla}. For the search for DM with mass around $10 \ \rm \mu eV$, the signal accumulation time can be up to $\simeq 0.1-1 \ \rm ms$ determined by the DM coherence time scale. There, for the qubit number $L\lesssim 10^{3}$, we anticipate that the $W$-state projection time is negligible compared to the signal accumulation time; beyond $L\sim 10^{3}$, the $W$ state projection time might be non-negligible, and the optimization of gate sequence or the shorter gate operation time is demanded. 
When probing lighter DM, the DM coherence time is longer, so the projection time could be less important.

Regarding the present performance of $W$ state projection, we comment that the deterministic algorithms to prepare the $W$ state with a general number of qubits within logarithmic steps are proposed in Ref.~\cite{Cruz_2019}, and in the work, they have already experimentally achieved high and fairly good fidelity of generating $W$ state with $L \sim 10$ using superconducting qubits on the IBM quantum computer. 
The inverse operation can be directly applied to project the sensors in the $W$ state in our protocol.

Another technical challenge is to align the qubit frequencies close together within the DM search bandwidth in order to realize the optimal signal.
If the frequencies of qubits differ significantly beyond the bandwidth of the measurement, $\Delta \omega_{\rm BW}$, the collective excitation to the $W$ state by the DM signal is suppressed, and the advantage of our protocol is reduced.
Qubit platforms with frequency tunability are particularly favorable, such as transmon superconducting qubits equipped with SQUID loop~\cite{Koch:2007hay} or NV center in diamond~\cite{Doherty_2013}, and routine frequency calibration prior to each measurement cycle is required.

Thus, the realistic step toward experimental implementation would be to realize the high-fidelity $W$ state projection with a large number of qubits beyond $\sim$10 and on experimental control to align the frequency of multiple qubits closely.
Based on the rapid development of quantum technology, including quantum metrology and quantum computing, we anticipate that such improvements could be achieved in the near future.

The key point of our protocol is that the DM signal collectively excites the qubits from the ground state to the $W$ state, while the noise and background mainly change the system's state to a state space perpendicular to that spanned by the ground and $W$ states.
In other words, the subspace spanned by the ground and $W$ states is the ``signal'' subspace, where the sensor state evolves under the effect of DM, while the noise effects mainly contribute to the change of qubits to other subspaces.
This observation suggests that one can extend our protocol to one similar to the quantum sensing with error correction~\cite{kessler2014quantum,dur2014improved,arrad2014increasing}, where one performs the error correction operations during the signal accumulation to remove the effect of some specific types of errors and project the state back to the signal subspace.
In our case of the DM search, the noise effect also induces the excitation of qubits within the signal subspace; we cannot completely remove the effect of noise by the error correction. However, we expect that the error correction can help mitigate the effect of background excitation\,\cite{FMS_inprep}.

Before closing this paper, let us discuss the potential extensions.
One qubit state, $\ket{0}$ and $\ket{1}$, can be regarded as the spin-1/2 state, each of which has $J_z=-1/2$ and $J_z=+1/2$, respectively, where $J_z$ is the $z$-component of the angular momentum operator $\mathbf{J}$. 
Then, $|0\rangle^{\otimes L}$ and $|W\rangle$ can be identified as the collective spin states with total angular momentum $J=L/2$ and $z$-component $J_z=-L/2$ and $J_z=-L/2+1$, respectively, which are called Dicke states~\cite{Dicke:1954zz}.
One may also consider the sensing protocols using other collective spin states with different $J$ and $J_z$ values.
In particular, the Dicke state with $J_z=0$ may be interesting since it enhances the signal amplitude\,\cite{Saleem:2023vse}.
However, in our setup of the DM search, the noise effects are not suppressed in this state, and the enhancement similar to the $W$ state case may not be expected.
    We leave this to future work.

\vspace{2mm}
\noindent
\acknowledgments
The work of TS was supported by the JSPS fellowship Grant No.\ 23KJ0678.
This work was supported by JSPS KAKENHI Grant Nos.\ 24K17042 [HF], 25H00638 [HF], and 23K22486 [TM].
In this research work, HF and TM used the UTokyo Azure (\url{https://utelecon.adm.u-tokyo.ac.jp/en/research_computing/utokyo_azure/}). 
HF would like to thank Akito Kusaka and Bin Xu for useful discussions.
MN would like to thank Matthew Steggles, Samantha Stever, and Juha Vartiainen for their continued interest in this research.

\appendix

\section{Review of the time evolution for the open quantum system}
\label{sec:appendix_Lindblad}
Here, we briefly review the time evolution of the open quantum system described by the Lindblad equation.
This appendix is based on Refs.\,\cite{Nielsen:2012yss,PreskillPh229}.
Let us consider a system whose state is described by a density matrix $\rho$.
Any effect on the system must be described by a map $\mathcal{E}$ on the density matrix, i.e., $\rho \rightarrow \mathcal{E}(\rho)$, which is called a quantum operation.
Since the density matrix after the map must still be a density matrix, the map $\mathcal{E}$ must be a trace-preserving and completely positive map. Moreover, if the initial state is a mixture of states, the map must linearly act on each component of the mixture.
If we require these properties, the map $\mathcal{E}$ can be expressed in the Kraus representation as
\begin{equation}
    \mathcal{E}(\rho)=\sum_i E_i \rho E_i^\dagger,
\end{equation}
where $E_i$ are operators satisfying the trace-preserving relation $\sum_i E_i^\dagger E_i=I$\,\cite{Nielsen:2012yss}.
The operators $E_i$ are called Kraus operators.
Physically, the map $\mathcal{E}$ describes the evolution of the system as a probabilistic mixture of the states $E_i \rho E_i^\dagger$, where each outcome occurs with probability $\text{Tr}(E_i \rho E_i^\dagger)$. In other words, after the quantum operation, the system is in the state $E_i \rho E_i^\dagger / \text{Tr}(E_i \rho E_i^\dagger)$ with probability $\text{Tr}(E_i \rho E_i^\dagger)$, and the overall state is given by the weighted sum over all possible outcomes.

The time evolution of the density matrix can also be described by the quantum operation.
Let us consider the time evolution of the density matrix from $t$ to $t+\Delta t$.
We assume that the time evolution is \emph{Markovian}, i.e., the time evolution depends only on the state at time $t$ and not on the history of the state before $t$.
Then, the time evolution can be described by a quantum operation $\mathcal{E}_{\Delta t}$ as
\begin{align}
    \rho(t+\Delta t)&=\mathcal{E}_{\Delta t}(\rho(t))\\
    &=\sum_i E_i(\Delta t) \rho(t) E_i^\dagger(\Delta t).    
\end{align}
Let us assume that the time evolution is continuous and differentiable.
Then, we can expand the Kraus operators for small $\Delta t$ as\,\cite{PreskillPh229}
\begin{gather}
    E_0(\Delta t)=I+(-iH+K)\Delta t,\\
    E_i(\Delta t)=\sqrt{\Delta t} L_i \quad (i\geq 1), \label{eq:lindblad_op_def}
\end{gather}
where $H$ and $K$ are Hermitian operators, and $L_i$ are operators.
The trace-preserving relation $\sum_i E_i^\dagger E_i=I$ leads to
\begin{align}
    K&=-\frac{1}{2}\sum_{i\geq 1} L_i^\dagger L_i.
\end{align}
Then, the time evolution of the density matrix is given by
\begin{align}
    \rho(t+\Delta t)&=\rho(t)+\Delta t \left[-i[H,\rho(t)]+\sum_{i\geq 1} \qty(L_i \rho(t) L_i^\dagger -\frac{1}{2}\{L_i^\dagger L_i,\rho(t)\}) \right]+O(\Delta t^2).
\end{align}
Taking the limit $\Delta t \rightarrow 0$, we obtain the equation
\begin{equation}
    \frac{d\rho}{dt}=-i[H,\rho]+\sum_{i\geq 1} D_i[\rho], \label{eq:Lindblad}
\end{equation}
where
\begin{equation}
    D_i[\rho]=L_i \rho L_i^\dagger -\frac{1}{2}\{L_i^\dagger L_i,\rho\}.
\end{equation}
This is called the Lindblad equation\,\cite{Lindblad:1975ef,Gorini:1975nb}.
The first term in the right-hand side of Eq.\,\eqref{eq:Lindblad} describes the unitary time evolution generated by the Hamiltonian $H$.
The other terms describe the non-unitary time evolution due to the environmental noise.
The operators $L_i$ are called Lindblad operators or jump operators.
By the deriviation, Eq.\,\eqref{eq:lindblad_op_def}, the Lindblad operator $L_i$ maps the state $\rho$ to $L_i \rho L_i^\dagger / \text{Tr}(L_i \rho L_i^\dagger)$ with the rate $\text{Tr}(L_i \rho L_i^\dagger) \Delta t$ in the infinitesimal time interval $\Delta t$.

\section{Solving the Lindblad equation with perturbation theory}
\label{sec:appendix_perturbation}
In this Appendix, we solve the Lindblad equation, Eq.~\eqref{eq: EOMc}, with the perturbation theory in terms of the signal strength $\epsilon$.
First, let us rewrite the equation of motion as
\begin{equation}
    \frac{d \vec{c}}{dt}=\qty[L_0+ L_1(t)] \vec{c},
\end{equation}
where $\vec{c}=\{c_{00},c_{01},c_{10},c_{11}\}$ and 
\begin{align}
    L_0 &=
    \begin{pmatrix}
    -\Gamma_0 & 0 & 0 & \Gamma_1 \\
    0 & -\gamma_2 & 0 & 0 \\
    0 & 0 & -\gamma_2 & 0 \\
    \Gamma_0 & 0 & 0 & -\Gamma_1
    \end{pmatrix}, \\
    L_1(t) &= \epsilon
    \begin{pmatrix}
    0 & i e^{-i \Delta \omega t} & -i e^{i \Delta \omega t} & 0 \\
    i e^{i \Delta \omega t} & 0 & 0 & -i e^{i \Delta \omega t} \\
    -i e^{-i \Delta \omega t} & 0 & 0 & i e^{-i \Delta \omega t} \\
    0 & -i e^{-i \Delta \omega t} & i e^{i \Delta \omega t} & 0
    \end{pmatrix}.
\end{align}
Here, we separate the Lindblad superoperator into two parts, $L_0$ and $L_1(t)$, where we take all the decoherence effects into account in $L_0$ and the signal effect in $L_1(t)$.
As mentioned in the main text, we consider the situation where the signal strength $\epsilon$ is much smaller than the decoherence effects $\Gamma_{0,1,2}$. Therefore, we treat $L_1(t)$ as a perturbation and solve the equation perturbatively.
First, we move to the interaction picture with respect to $L_0$ by defining a new parameter $\vec{C}(t)$ as
\begin{equation}
    \vec{C} (t) = e^{-L_0 t} \vec{c}(t). \label{eq:intpicC}
\end{equation}
Using this parameter, the equation of motion is rewritten as
\begin{equation}
    \frac{d \vec{C}}{dt}= L_I(t)\vec{C},
\end{equation}
where
\begin{equation}
    L_I(t)=e^{-L_0 t} L_1(t) e^{L_0 t}.
\end{equation}
Then, the solution can be solved perturbatively with respect to $L_I$. Since we consider a qubit initialized in the ground state, $\vec{C}_0=\vec{C}(0)=\{1,0,0,0\}$, we obtain the solution at time $t$ as
\begin{equation}
    \vec{C}(t) 
    = \vec{C}_0 + \int_0^t dt_1 L_I(t_1) \vec{C}_0 + \int_0^t dt_1 \int_0^{t_1} dt_2 L_I(t_1) L_I(t_2) \vec{C}_0 + O(\epsilon^3). \label{eq:perturbative_solution}
\end{equation}
The original vector $\vec{c}(t)$ can be recovered by multiplying $e^{L_0t}$ to Eq.~\eqref{eq:perturbative_solution} according to Eq.~\eqref{eq:intpicC}.

\section{Uncertainty of the parameter estimation}
\label{sec:appendix_parameter_estimation}
In this appendix, we briefly review the uncertainty of the parameter estimation in quantum metrology.
We discuss the relation between the signal-to-noise ratio and the uncertainty of the parameter estimation.

Suppose that we want to estimate a single parameter $\epsilon$ by measuring an observable $A$. For simplicity, we assume that each measurement of $A$ gives either $0$ or $1$; i.e., $A$ is a projection operator such as $(\sigma_z+I)/2$ for a qubit. If we repeat the measurement $N$ times, the estimator of $A$, denoted by $\hat{A}$, is given by
\begin{align}
    \hat{A} &= \frac{1}{N} \sum_{i=1}^N A_i,
\end{align}
where $A_i$ is the outcome of the $i$-th measurement.
For large $N$, the estimator $\hat{A}$ converges to the quantum mechanical expectation value, $\langle A \rangle = \text{Tr}(\rho A)$, where $\rho$ is the density matrix of the system.
The parameter $\epsilon$ is estimated as a function of $\hat{A}$ as $\hat{\epsilon} = \hat{\epsilon}(\hat{A})$.

We are now interested in the uncertainty of the estimator $\hat{\epsilon}$.
Before discussing the uncertainty of $\hat{\epsilon}$, let us first consider the uncertainty of $\hat{A}$. Since each measurement of $A$ gives either $0$ or $1$, $A_i$ follows the Bernoulli distribution with the probability $p=\langle A \rangle$. 
Then, the sum, $N \hat{A}$, follows the binomial distribution with the mean $Np$ and the variance $Np(1-p)$. Namely, the standard deviation of $\hat{A}$ is estimated as
\begin{align}
    \delta \langle A \rangle &= \sqrt{\frac{p(1-p)}{N}} = \frac{\sqrt{\left\langle \qty(A - \langle A \rangle)^2 \right\rangle}}{\sqrt{N}}.
\end{align}
For large $N$, the standard deviation $\delta \langle A \rangle$ is small enough and we may estimate the standard deviation of $\epsilon$ by using the linear approximation as
\begin{align}
    \delta \epsilon &= \frac{\delta \langle A \rangle}{\abs{d\langle A \rangle/d\epsilon}} = \frac{\sqrt{\left\langle \qty(A - \langle A \rangle)^2 \right\rangle}}{\sqrt{N}\abs{d\langle A \rangle/d\epsilon}}.
\end{align}

In usual situations for the DM search, we would like to exclude some parameter region of $\epsilon$, e.g., $\epsilon > \epsilon_0$, at a certain confidence level, based on the measurement result, which typically gives a small value of $\hat{A}_0$ consistent with $\epsilon=0$.
In this case, the null hypothesis is $\epsilon=\epsilon_0$ and we would like to check whether the measurement result, $\hat{A}_0$, is consistent with the null hypothesis.
Since $N\hat{A}$ follows the binomial distribution, we may perform the likelihood-ratio test to check the significance of the estimation of $\epsilon$ for $\hat{A}$. In particular, if $N$ is large enough and $N p$ is not too small, the distribution of $N\hat{A}$ can be approximated by the normal distribution with the mean $N p$ and the variance $N p(1-p)$.
Then, the standard score $z$ is given by
\begin{align}
    z &= \left|\frac{\langle A \rangle|_{\epsilon=\epsilon_0} - \hat{A}_{0}}{\delta \langle A \rangle}\right| \simeq \left|\frac{\langle A \rangle|_{\epsilon=\epsilon_0} - \langle A \rangle|_{\epsilon=0}}{\sqrt{\langle A \rangle|_{\epsilon=\epsilon_0}/N}}\right|, \label{eq:standard_score}
\end{align}
where we assume that $\langle A \rangle|_{\epsilon=\epsilon_0}$ is small enough. Here, instead of the real measurement result $\hat{A}_0$, we approximate it by $\langle A \rangle|_{\epsilon=0}$ to discuss the expected sensitivity of the experiment.
If, for example, $z>1.96$, we may reject the null hypothesis at the 95\% confidence level and conclude that $\epsilon > \epsilon_0$ is excluded at the 95\% confidence level.
If we may approximate $\langle A \rangle|_{\epsilon=\epsilon_0} - \langle A \rangle|_{\epsilon=0} \simeq (d\langle A \rangle/d\epsilon|_{\epsilon=\epsilon_0}) \epsilon_0$, the standard score $z$ is rewritten as
\begin{align}
    z &\simeq \left.\frac{\epsilon}{\delta \epsilon}\right|_{\epsilon=\epsilon_0}.
\end{align}

This standard score $z$ is similar to the signal-to-noise ratio, although it corresponds to the discovery search and the null hypothesis is rather $\epsilon=0$.
To see this, let us separate the expectation value $\langle A \rangle$ into the background and signal contributions. Namely, we write $\langle A \rangle$ as
\begin{align}
    \langle A \rangle &= \langle A \rangle_{\rm BG} + \langle A \rangle_{\rm sig}(\epsilon),
\end{align}
where $\langle A \rangle_{\rm BG}$ and $\langle A \rangle_{\rm sig}(\epsilon)$ are the background and signal contributions, respectively. We also assume that $\langle A \rangle_{\rm sig}(\epsilon)$ is small enough compared to $\langle A \rangle_{\rm BG}$, although $\langle A \rangle_{\rm BG}$ itself is small as well. If we repeat the measurement $N$ times, the ``signal'' $S$ and the ``background'' $B$ are given as
\begin{align}
    S &= N \langle A \rangle_{\rm sig},\\
    B &= N \langle A \rangle_{\rm BG}.
\end{align}
The
signal-to-noise ratio, $S/\sqrt{B}$, is given by
\begin{align}
    \frac{S}{\sqrt{B}} &= \frac{\sqrt{N} \langle A \rangle_{\rm sig}}{\sqrt{\langle A \rangle_{\rm BG}}}.
\end{align}

\section{Quantum Cramér-Rao bound}
\label{app:QCRB}
Here, we estimate the theoretical bound of the uncertainty of the estimation of $\epsilon$ when one starts from all qubits in the ground state and let them evolve independently, using the quantum Cramér-Rao bound\,\cite{Paris:2008zgg}.
The quantum Cramér-Rao bound states that, when one tries to estimate the parameter $\epsilon$ from quantum state $\rho$, the variance of the parameter estimation is bounded by:
\begin{equation}
    (\delta \epsilon)^2 \gtrsim \frac{1}{N F(\theta)},
\end{equation}
where $N=T/t$ is the total number of  measurements each with time $t$ within fixed total time $T$, and  $F(\epsilon)$ is the quantum Fisher information of a state $\rho$. With the decomposition of state $\rho$  as 
$\rho=\sum_a \Lambda_a \ketbra{\Psi_a}$, 
the quantum Fisher information $F(\epsilon)$ is given by
\begin{equation}
    F(\epsilon) = 2\sum_{a,b} \frac{| \bra{\Psi_a} \partial_\epsilon  \rho \ket{\Psi_b}|^2}{\Lambda_a+\Lambda_b}.
\end{equation}

To find the quantum Cramér-Rao bound for our situation when we have $L$ identical qubits evolve independently, first, we show that the quantum Fisher information of the tensor product of $L$ identical qubits is simply $L$ times that of a single qubit.
Then, the bound can be calculated from the quantum Fisher information of a single qubit.

 In our protocol, we do not consider the entanglement between qubits, and the entire state can be simply written as a tensor product of each qubit's density matrix as
\begin{align}
    \rho(t)=& \rho_1(t)\otimes \rho_2(t) \otimes... \otimes \rho_L(t),
\end{align}
where each qubit density matrix is treated to be identical to the others.
In general, a qubit density matrix can be obtained by spectral decomposition and written as
\begin{equation}
    \rho_i
    =
    \lambda_0 \ketbra{\psi_0}+\lambda_1 \ketbra{\psi_1}
\end{equation}
with $\langle \psi_a|\psi_b\rangle=\delta_{a,b}$ and $\lambda_0+\lambda_1=1$.
Then, the density matrix of the total system can be decomposed as
\begin{align}
    \rho(t)&=\sum_{\{x_i=0,1\}}\lambda_0^{L-\sum_ix_i}\lambda_1^{\sum_ix_i}\ketbra{\psi_{x_1}} \otimes  \ketbra{\psi_{x_2}}\cdots\otimes \ketbra{\psi_{x_L}}\nonumber \\
    &=\sum_{\vec{x}}\Lambda_{\vec{x}}\ketbra{\Psi_{\vec{x}}}
\end{align}
with $\vec{x}\equiv\{x_1,x_2,\dots,x_L\}$, and
\begin{gather} 
    \Lambda_{\vec{x}}=\lambda_0^{L-\sum_ix_i}\lambda_1^{\sum_ix_i}, \quad \ket{\Psi_{\vec{x}}}\equiv \ket{\psi_{x_1}}\otimes \ket{\psi_{x_2}}\otimes... \otimes \ket{\psi_{x_L}}.
\end{gather}

The Fisher information can then be calculated by
\begin{align}
    F(\epsilon)= 2\sum_k\sum_{\vec{x}_i,\vec{x}_j}\frac{|\bra{\Psi_{\vec{x}_i}}\rho_1  ...  \otimes\frac{d\rho_k}{d\epsilon}... \otimes\rho_L \ket{\Psi_{\vec{x}_j}}|^2}{\Lambda_{\vec{x}_i}+\Lambda_{\vec{x}_j}}.
\end{align}
Since only the entry that is not differentiated by $\epsilon$ is diagonal, i.e., $\bra{\psi_a} \rho_i \ket{\psi_b} =\lambda_{a} \delta_{a,b}$, one can further simplify $F(\epsilon)$ to
\begin{align}
    F(\epsilon)
    &=
    2\sum_k
    \sum_{\{x_i\}_{i\neq k}}\sum_{a,b}\lambda_0^{L-\sum_{i\neq k}x_i}
    \lambda_1^{\sum_{i\neq k}x_i} \frac{ 
    |   
    \bra{\psi_a} d\rho_k/d\epsilon \ket{\psi_b}|^2}{ (\lambda_a+\lambda_b)}\nonumber \\
     &=2\sum_k \sum_{a,b} 
     \frac{|\langle \psi_{a}| d\rho_k /d\epsilon|\psi_{b} \rangle|^2}{\lambda_{a}+\lambda_{b}}\nonumber \\
     &=L\left( 2\sum_{a,b} 
     \frac{|\langle \psi_{a}| d\rho_1 /d\epsilon|\psi_{b} \rangle|^2}{\lambda_{a}+\lambda_{b}}\right)
\end{align}
where in the second equation we used that 
\begin{equation}
    \sum_{\{x_i=0,1\}_{i\neq k}} \lambda_0^{L-\sum_{i\neq k}x_i}
    \lambda_1^{\sum_{i\neq k}x_i}=(\lambda_0+\lambda_1)^{L-1}=1
\end{equation}
to sum over all possible values of $x_i$ with $i\neq k$,
and, in the last line, we used that all the qubit density matrices are identical. The last line shows that the total Fisher information is simply $L$ times that of a single qubit.

Now we calculate the quantum Fisher information of a single qubit:
\begin{equation}
    F^{(1)}(\epsilon)
    =2\sum_{a,b} 
     \frac{|\langle \psi_{a}| d\rho_1 /d\epsilon|\psi_{b} \rangle|^2}{\lambda_{a}+\lambda_{b}}.
\end{equation}
The solution of one qubit density matrix $\rho_1$ governed by Eq.~\eqref{eq: EOMc} is shown in Eqs.~\eqref{eq:c01}, \eqref{eq:c10} and \eqref{eq:c11}.
Based on them, at leading order, we obtain that
\begin{equation}
    \frac{d \rho_1}{d\epsilon}=\frac{1}{\epsilon} 
    (c_{01} \ketbra{0}{1}
    +c_{10} \ketbra{1}{0})+ O(\epsilon).
\end{equation}
With $\epsilon$ as a perturbation, we also know that the density matrix $\rho_1$ can be decomposed by
\begin{align}
    \ket{\psi_0}&=\ket{0}+O(\epsilon)\\
    \ket{\psi_1}&=\ket{1}+O(\epsilon),
\end{align}
with $\lambda_0 = 1-p_{1,\rm BG}+O(\epsilon^2)$ and $\lambda_1 = p_{1,\rm BG}+O(\epsilon^2)$.
This gives 
\begin{equation}
    F^{(1)} = 4  |c_{01}|^2/\epsilon^2+O(\epsilon).
\end{equation}
Putting this into the quantum Fisher information of the total system, we have
\begin{equation}
    F(\epsilon)= L\frac{4|c_{01}|^2}{\epsilon^2} + O(\epsilon)
\end{equation}
giving the Quantum Cramér-Rao bound as
\begin{equation}
    (\delta \epsilon)^2 \gtrsim \frac{1}{L}\frac{t}{T}\frac{\epsilon^2}{|c_{01}(t)|^2}.
\end{equation}
Optimization of the time $t$ for each measurement to minimize the bound gives
\begin{equation}
    \delta \epsilon \gtrsim \frac{1}{\sqrt{L}}\sqrt{\frac{\gamma_2}{T} }.
\end{equation}
Note that the uncertainty $\delta \epsilon^{(W)}$ with our protocol that projects sensors to $W$ state, Eq.~\eqref{eq:depW}, reproduces this limit when $L$ is so large that the number of signals during the measurement time $t\sim 1/\gamma_2$ is comparable to the background excitation, i.e.,
\begin{equation}
    L\frac{\epsilon^2}{\gamma_2} \sim \Gamma_0,
\end{equation}
while satisfying the condition $L\Gamma_0/\gamma_2 \lesssim 1$.

\bibliographystyle{JHEP.bst}
\bibliography{papers}

\end{document}